\documentclass[%
 reprint,
 amsmath,amssymb,
 aps,
]{revtex4-1}

\usepackage{graphicx}
\usepackage{dcolumn}
\usepackage{bm}

\begin{document}


\title{Work function measurements of vanadium doped diamond-like carbon films by ultraviolet photoelectron spectroscopy }

\author{Akihiko Shigemoto}
 \email{shige@wakayama-kg.go.jp}
\affiliation{Industrial Technology Center of Wakayama Prefecture}

\author{Tomoko Amano}
\author{Ryozo Yamamoto}
\affiliation{SEAVAC Inc.}

\date{\today}

\begin{abstract}
Vanadium doped diamond-like carbon films prepared by unbalanced magnetron sputtering have been investigated by X-ray and ultraviolet photoelectron 
spectroscopy measurements for the purpose of revealing electronic structures including values of work function on the surfaces. In addition to these 
photoelectron measurements, X-ray diffraction measurements have been performed to characterize the crystal structures.
\end{abstract}

\maketitle


\section{Introduction}
Vacuum deposition is a powerful tool to coat films which compose of high melting point materials, such as amorphous carbon, metallic carbides and so on. Among these materials, an amorphous carbon film, which is industrially called diamond-like carbon (DLC), is attractive to a widespread use of protective coatings from hard disk drives to automobile engines. DLC compose of sp$^{2}$-, sp$^{3}$- orbital configurations and hydrogen atoms and the sp$^{3}$-configuration is denoted as $\sigma $-bonding as in diamond, which forms a tetrahedral structure. In the sp$^{2}$-configuration as in graphite, 
three of four valence electrons are denoted as $\sigma $-bonding in plane and the other one electron occupies $\pi $-orbital \cite{Aisenberg, Ferrari, Robertson1, Robertson2}. 
Because of diamond structure, DLC surface shows hardness up to 80 GPa \cite{McKenzie, Fallon, Pharr}. 
In order to improve mechanical properties and coating adhesions, transition metals are utilized as additives for DLC coatings. 
For instance, it is known that vanadium carbide coatings have great abrasion resistance \cite{Paton, Aouni, Sainte}.

In this study, vanadium doped DLC films have been prepared by unbalanced magnetron sputtering (UBMS) and analyzed by X-ray and ultraviolet photoelectron spectroscopy measurements (XPS and UPS) to reveal electronic structures of the films. Moreover, X-ray diffraction (XRD) measurements have been performed in order to characterize the crystal structures.


\begin{figure}
\includegraphics[width=\columnwidth]{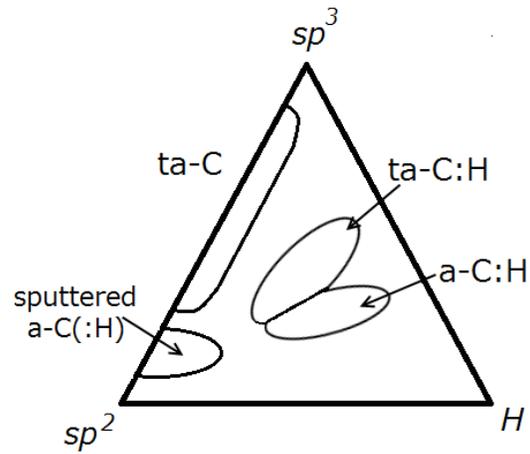}
\caption{\label{DLCdiagram.eps} Schematic Diagram of DLC (The original one is denoted in Ref.\cite{Robertson1}.)}
\end{figure}

\section{Experimental}
Two vanadium doped DLC films (sample A and B) were deposited on single crystalline silicon substrates (100) by unbalanced magnetron sputtering 
(UBMS) \cite{UBMS1, UBMS2}. In UBMS process, magnets are set for the purpose of promoting an ionization of sputtered atoms. The sample substrates 
were heated up to 473 K and sputtered for 180 minutes with -50 V bias voltage. Vacuum pressure has been kept 0.5 Pa during the coating process. 
The difference of these two sample were sputtering energies on vanadium and carbon targets. 
In the case of sample A, the sputtering energies of vanadium and carbon targets were 1.5 kW and 0.5 kW respectively and the sputtering energies on both vanadium and carbon targets are 1.5 kW on sample B. 

The values of surface hardness on vanadium doped DLC films have been evaluated by micro hardness testing machine Shimadzu DUH-200 \cite{ DUH}.

On photoelectron spectroscopy measurements, the light sources of X-ray and ultraviolet photoelectron spectroscopies (XPS, UPS) were monochromatic Al K$\alpha $ (h$\nu $ = 1486.6 eV) and He I discharge (h$\nu $ = 21.2 eV) with a hemisphere electron analyzer JEOL JPS9010-MC. The sample surfaces were cleaned by Ar$^{+}$ sputtering in the preparation chamber and a series of spectra have been obtained under 2 $\times$ 10$^{-7}$ Pa. 
The full width of the half maximum (FWHM) of Ag 3d$_{7/2}$ peaks was 0.6 eV on XPS measurements with monochrome meter and the energy resolution of UPS was estimated as 85 meV by spectrum fitting on Au Fermi edge at room temperature. In order to characterize crystal structures of the samples, XRD patterns have been measured by excitation of Cu K$\alpha $ radiation (the wavelength is 1.54 
{\AA}). With use of the experimental apparatus of RIGAKU RINT 1400, the 2 $\theta$ scan swept from 30 to 100 $^\circ$ by 0.02 $^\circ$ sampling step.

\section{Results and Discussion}

\begin{figure}
\includegraphics[width=\columnwidth]{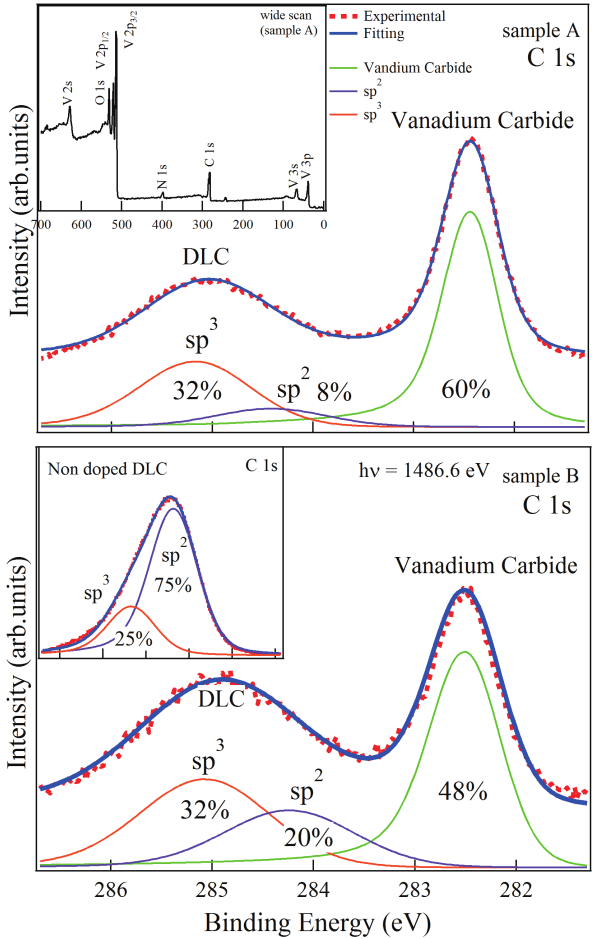}
\caption{\label{XPS_C1s.eps} XPS C1s peaks of vanadium doped DLC films. (sample A: V(1.5kW), C(0.5 kW) sample B: V(1.5kW), C(1.5 kW) )}
\end{figure}

The thickness of sample A was estimated as 0.9 $\mu $m according to the measurement of the step between the DLC surface and the silicon substrate. 
With 0.2 gf (1.96 $\times$ 10$^{-3}$ N) loading, dynamic micro hardness tests have also been performed and the surface hardness showed up to 3000 and 2350 DHV (Dynamic Vickers Hardness) on sample A and B respectively in comparison with 1950 DHV hardness of the non doped DLC film.

\begin{figure}
\includegraphics[width=\columnwidth]{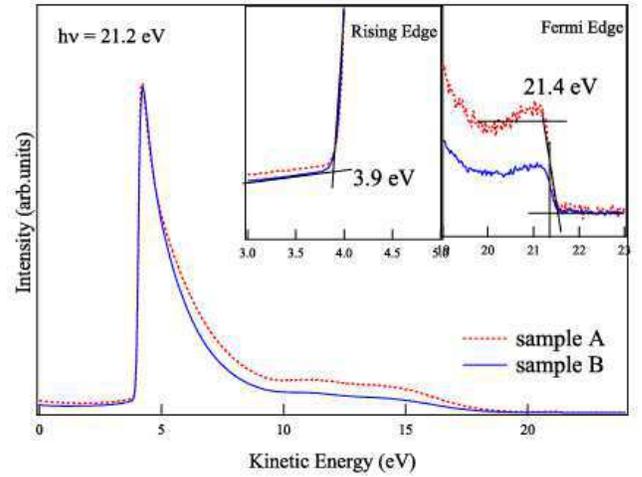}
\caption{\label{UPS_DLC.eps} Work function measurements of vanadium doped DLC surfaces by UPS.}
\end{figure}

\begin{figure}
\includegraphics[width=\columnwidth]{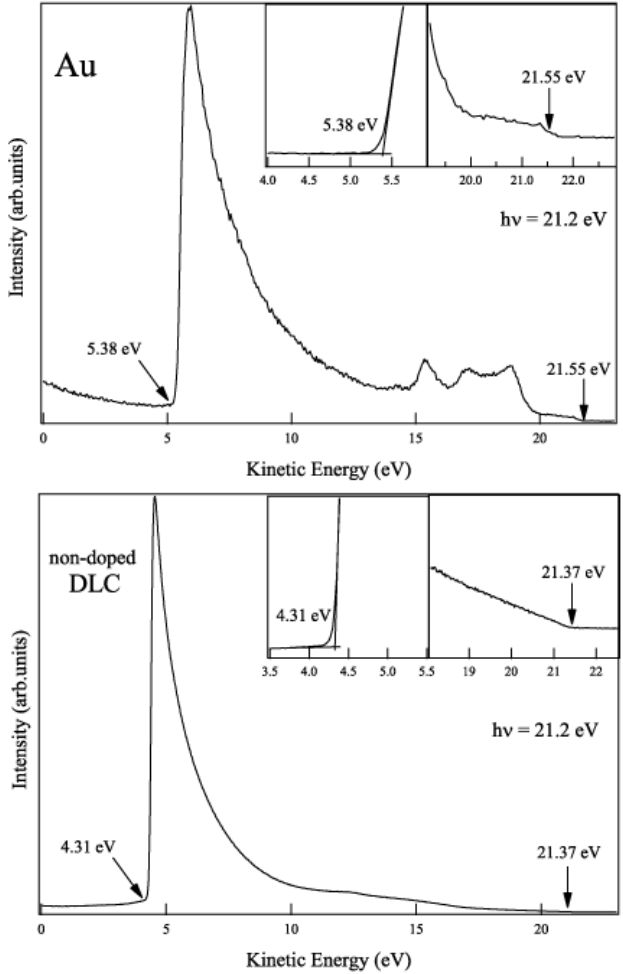}
\caption{\label{UPS_Au.eps} Work function measurements of Au and non doped DLC surfaces by UPS}
\end{figure}

In Fig.\ref{XPS_C1s.eps}, the C1s peaks of both sample A and B have been split into two components, 
these binding energies are 285 eV and 282.5 eV respectively. One binding energy 285 eV is typical of a DLC film 
and includes both carbon sp$^{2}$ and sp$^{3}$ hybridized components. 
The other binding energy 282.5 eV should be assigned to carbon 1s components which derives from vanadium carbide. 
In the case of photoelectron spectroscopy on vanadium carbide V$_{8}$C$_{7 }$, the carbon 1s peak is observed on 282.5 eV which binding energy is lower than the graphite peak \cite{Choi}. By the spectral weights estimation, 
the component ratios of vanadium carbide are 60{\%} on sample A and 48 {\%} on sample B respectively.
As for C1s DLC peaks, they should be divided into sp$^{2}$ and sp$^{3}$ components. 
On sample A, spectral weights of sp$^{2}$ and sp$^{3}$ components are 8 {\%} and 32 {\%}, in other words, 
the relative ratio of sp$^{2}$: sp$^{3}$ is 20 : 80. Similarly the spectral weights of sp$^{2}$ and sp$^{3}$ components on sample B are 20 {\%} and 32 {\%}, therefore the relative ratio of sp$^{2}$ : sp$^{3}$ is 38 : 62. 
Considering the relative ratio of sp$^{2 }$: sp$^{3}$ components on non doped DLC is 75:25, 
the surface hardness increases with increasing sp$^{3}$ components.

\begin{figure}
\includegraphics[width=\columnwidth]{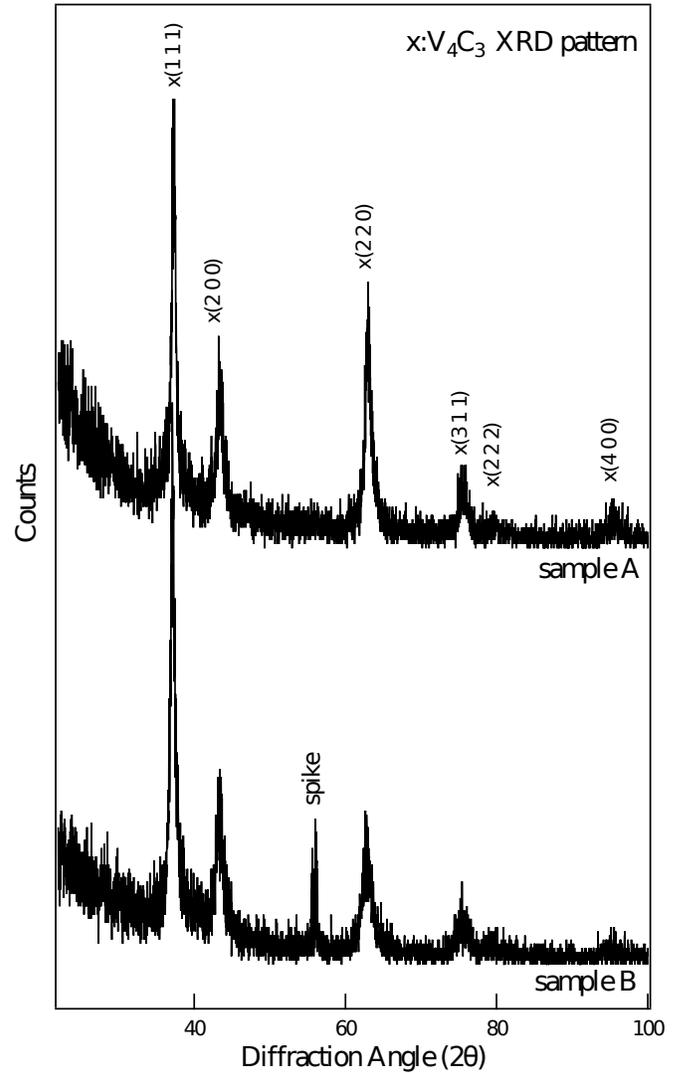}
\caption{\label{XRD.eps} XRD pattern of vanadium doped DLC. The indexes of diffraction pattern are given by CAS $\backslash ${\#} 12070-10-9.}
\end{figure}


Furthermore, we have measured UPS as shown in Fig.\ref{UPS_DLC.eps} and the kinetic energy of the electrons at the Fermi edge (E$_{Fermi})$ and that of the secondary electron cutoff (E$_{cutoff})$ are enlarged at the right shoulder of Fig.\ref{UPS_DLC.eps}. To distinguish secondary electron cutoff from the low kinetic energy electron scattering, Negative bias voltages were put on the sample. 
Plural negative bias voltages (-8, -10, -12 V) were selected to check a charge on the sample surface. 
The 10 V bias was employed in the work function measurements, however the horizontal axes of the spectra in Fig.\ref{UPS_DLC.eps} and \ref{UPS_Au.eps} have been corrected by 10 eV towards a positive direction \cite{Smith, Barnard, Bondzie, Yoshitake1,Yoshitake2}. Concerning sample A shown in Fig.\ref{UPS_DLC.eps}, the binding energy E$_{Fermi}$ and the E$_{cut-off}$ are 21.4 and 3.9 eV respectively, 
therefore a work function $\Phi $ is estimated as below 

$\Phi \quad =$ h$\nu $ - (E$_{Fermi}$ -- E$_{cut-off})$
\par = 21.2 -- (21.4 -- 3.9) 
\par= 3.7 eV

In the case of sample B, the work function was also estimated as 3.7 eV same as that of sample A. In addition to these samples, we have also measured UPS of gold plate and non doped DLC in order to verify the correctness of a series of measurements. The obtained values were 5.0 eV on gold and 4.1 eV on DLC.
The work function of Au corresponds to the reference.\cite{Eastman} In the case of DLC, the atomic structure depends 
on hybridization of carbon sp$^{2}$ and sp$^{3}$ components, therefore the work function value is not unique 
and it was reported the value is from 4 to 5 eV.\cite{Robertson1} The estimated value 4.1 eV was within this range.

In order to characterize the crystal structures of the samples, XRD measurements have been obtained. 
The XRD patterns of sample A and B are illustrated in Fig.\ref{XRD.eps}. 
The main peaks of these two samples correspond with the XRD pattern of vanadium carbide V$_{4}$C$_{3}$ and V$_{8}$C$_{7}$. 
The diffraction points are (111), (200), (220), (311), (222) and (400).\cite{V4C3} 
On sample B, a spike is observed at 56 degree and it is not a signal but a hardware error. The diffraction pattern of 
V$_{8}$C$_{7}$ is similar to that of V$_{4}$C$_{3}$, however, V$_{8}$C$_{7}$ has additional diffraction points. 
Among these additional diffraction points, some large points are (520), (521), (720) and (650) which are located between 59.7 and 92.4 degree.\cite{V8C7} These additional diffraction points of V$_{8}$C$_{7}$ have not been observed, hence, the crystal structure on the surface should be assigned to V$_{4}$C$_{3}$. 

\section{Conclusion}
Vanadium doped DLC film prepared by unbalanced magnetron sputtering has been investigated by X-ray and ultraviolet 
photoelectron spectroscopies. The XPS spectrum has two C1s peaks, one binding energy is 285 and the other is 282.5 eV. These two components are 
assigned to DLC and vanadium carbide components respectively. In addtion to XPS measurements, the value of work function has been estimated as 3.7 eV by 
UPS.

\begin{acknowledgments}
We authors thank Mr. Kitayama (The Advanced Materials Processing Institute Kinki Japan) for manipulating the UBMS apparatus on the DLC coating process. 
\end{acknowledgments}

\bibliographystyle{prb}
\bibliography{references}

\end{document}